\documentclass[aoas]{imsart}
\RequirePackage{amsthm,amsmath,amsfonts,amssymb}
\RequirePackage[authoryear]{natbib}
\RequirePackage{graphicx}

\startlocaldefs
\def\journal@name{}
\endlocaldefs

\begin{document}

\begin{frontmatter}
\title{Tunable robustness in power-law inference}

\begin{aug}
\author[A,B]{\fnms{Qianying} \snm{Lin}\ead[label=e1]{qianying@lanl.gov}},
\and
\author[C,D]{\fnms{Mitchell} \snm{Newberry}\ead[label=e2]{mgnew@umich.edu}}
\address[A]{Theoretical Biology and Biophysics,
  Los Alamos National Laboratory, \printead{e1}}
\address[B]{Michigan Institute for Data Science,
  University of Michigan}
\address[C]{Department of Human Behavior, Ecology and Culture,
  Max Planck Institute for Evolutionary Anthropology}
\address[D]{Center for the Study of Complex Systems,
  University of Michigan, \printead{e2}}
\end{aug}

\begin{abstract}
Power-law probability distributions arise often in the
social and natural sciences. Statistics have been
developed for estimating the exponent parameter as well
as gauging goodness-of-fit to a power law. Yet
paradoxically, many famous power laws such as the
distribution of wealth and earthquake magnitudes have
not found good statistical support in data by modern
methods. We show that measurement errors such as
quantization and noise bias both maximum-likelihood
estimators and goodness-of-fit measures.  We address
this issue using logarithmic binning and the
corresponding discrete reference distribution for
maximum likelihood estimators and Kolmogorov-Smirnov
statistics. Using simulated errors, we validate that
binning attenuates bias in parameter estimates and
recalibrates goodness of fit to a power law by removing
small errors from consideration. These benefits come at
modest cost in statistical power, which can be
compensated with larger sample sizes. We reanalyse
three empirical cases of wealth, earthquake magnitudes
and wildfire area and show that binning reverses
statistical conclusions and aligns the statistical
results with historical and scientific expectations.
We explain through these cases how routine errors lead
to incorrect conclusions and the necessity for more
robust methods.
\end{abstract}

\end{frontmatter}

\section{Introduction} Power laws---where a quantity
$Y$ scales as a constant power of another quantity $X$
according to the form $Y \propto X^\alpha$---are
ubiquitous in nature and arise for a variety of reasons
\citep{reed2002gene,willinger2004more,newman2005power,gabaix2016power}.
In biology, power laws appear as allometric
relationships in physiology and morphology. Such
allometries classically established theoretical limits
on the heights of trees \citep{thompson1917growth} and
the weights of dinosaurs \citep{anderson1985long}. In
physics, power laws are a hallmark of self-similar and
scale-free systems, where the exponent $\alpha$
specifies how to translate from one scale to another.
Power laws have been observed or claimed in systems as
wide-ranging as skeletal morphology, terrorism, baby
names \citep{hahn2003drift} and urban infrastructure
\citep{bettencourt2013origins}, while the standards for
scientific or statistical justification have varied
widely between fields.

Probability distributions with power-law tails are a
special case where the probability of an observation
scales as a power of its magnitude, with the
consequence that extreme magnitude events are far more
likely than in the normal or exponential distributions.
In the case of a power-law distribution over a
continuous quantity $x$, both the tail distribution
function
\begin{equation}
\operatorname{Prob}(\mathrm{obs.} > x) = \left({x \over
x_m}\right)^{-\alpha}
\label{eq:tail}
\end{equation}
and the corresponding probability density function
\begin{equation}
p(x) = \alpha x_m^\alpha x^{-(\alpha + 1)}
\label{eq:dcpl}
\end{equation}
take the form of a power law above some threshold value
$x_m > 0$.  This continuous case is commonly called the
Pareto distribution, so named for Pareto's 1895
observation of power-law scaling in the frequencies of
extreme wealth and income \citep{pareto1895legge}.
Earthquake magnitudes were also classically observed to
follow a power-law distribution, which inspired the
ongoing practice of recording earthquake magnitudes on
the logarithmic Richter scale
\citep{richter1935instrumental,gutenberg1944frequency}.

Historically, regression on log-log plots was the means
to estimate the exponent $\alpha$ in power laws by
fitting the equation $(\log Y) = \alpha (\log X) + c$
to the data.  These methods have been equally applied
to probability distributions as to bivariate
relationships such as body mass and femur length.  For
example \citet{gutenberg1944frequency} used ordinary
least squares, taking $Y$ to be earthquake frequencies
within binned magnitude categories $X$, in order to fit
$\alpha$ using Eq.~\ref{eq:dcpl}. Different regression
models assume different error models for $X$ and $Y$
\citep{warton2006bivariate}. Ordinary least squares
assumes perfect knowledge of $X$ and identical normal
error on $Y$, whereas major axis regressions allow
errors on both $X$ and $Y$. None of these error models,
however, are exactly matched to samples from a
probability distribution \citep{bauke2007parameter}.

Samples do not have the same statistical variability as
errors on independent measurements, and parameters of
probability distributions have different constraints
than regression parameters. For example, probability
distributions require that $\int_{x_m}^\infty p(x)\, dx
= 1$, but using regression to infer the slope,
$-(\alpha + 1)$, and intercept, $\log \alpha
x_m^\alpha$, does not guarantee that this is the case
and typically produces a contradiction.
Regression methods are also considered potentially
problematic as there are few theoretical expectations
or guarantees about the accuracy or precision of the
resulting estimates
\citep{goldstein2004problems,bauke2007parameter}.
Furtheremore, residuals offer little useful information
about goodness of fit if the error models are
inappropriate to begin with.  Regression nonetheless
often happens to produce accurate estimates of $\alpha$
when applied to the empirical cumulative distribution
function, Eq.~\ref{eq:tail}, taking $X$ to be the
magnitudes in the data and $Y$ to be their quantiles.
This method remains popular in some fields and
continues to be improved \citep{gabaix2011rank}.

Another branch of well-developed theory treats the
general case of estimating parameters and gauging
goodness of fit given samples from a hypothesized
probability distribution. Parameter estimation using
maximum likelihood is guaranteed to be asymptotically
normal, unbiased, consistent, and optimally efficient
in the limit of large data
\citep{cramer1946mathematical} and corresponds to
maximum \textit{a posteriori} estimation in Bayesian
statistics \citep{jaynes2003probability}.
Maximum-likelihood estimators (MLEs) have been derived
specifically for power-law exponents
\citep{muniruzzaman1957measures,virkar2014power} and
correspond to the Hill index estimator in extremal
value theory \citep{hill1975simple}. Likewise, gauging
goodness of fit using Kolmogorov-Smirnov (K-S)
statistics is broadly applicable, firmly grounded in
probability theory, and easily adaptable to a
hypothesis testing framework
\citep{massey1951kolmogorov}. Goodness-of-fit tests
based on the K-S statistic have also been developed for
power-law data
\citep{goldstein2004problems,clauset2009power} and
binned power-law data \citep{virkar2014power}.

However, there are clues that MLEs and goodness-of-fit
tests based on K-S statistics give unreasonable answers
for power laws in practice. In fields such as
neuroscience and vascular biology, the goal is to
compare empirical estimates to sound theoretical
predictions, and here it is sometimes the estimates
that have noticeable flaws, leading empiricists to
question \citep{langlois2014maximum} or refuse
\citep{newberry2015testing} maximum likelihood methods.
For example, scaling exponents $\alpha$ for the
diameters of branching organs such as tree branches,
blood vessels and bronchia have a long-established
theoretical range between 2 to 3 as a direct
consequence of fluid mechanics
\citep{murray1926vascular,zamir1982arterial,west1997general},
yet $\alpha$ MLEs routinely fall well outside this
range even on ample high quality, hand-curated data
\citep{yeh1976tracheobronchial,newberry2019self}.
Goodness-of-fit tests using K-S statistics, on the
other hand, reject the power law in the otherwise
exemplary empirical cases of wealth and earthquake
magnitudes \citep{clauset2009power}.

Here we explain how such discrepancies between
statistical and scientific conclusions can be
attributed to inappropriate assumptions about error.
The theoretical justifications for MLEs and K-S
statistics assume that the hypothesized distribution of
the sample is known exactly, including any errors
associated with each data point. This assumption almost
never holds in practice: a hypothesized distribution is
only an approximation to the distribution of the
empirical data, because the real process of generating
scientific data incorporates known and unknown error
sources including measurement and recording errors.

We show that even an error of $\pm 0.2$ in a dataset
that spans a range from 1 to 100 is capable of biasing
the MLE by more than 10\%. Whereas in normal
statistics, a random measurement can be combined with
an unbiased error without affecting the shape of the
sample distribution or biasing estimates of the mean,
we show that in power-law distributions, even small
normal measurement errors qualitatively change the
shape of the sample distribution and bias estimates of
the slope $\alpha$. Small errors can then alter
conclusions of goodness-of-fit tests in common sample
sizes.

Unfortunately, the MLEs and K-S statistics developed
for power laws fail to achieve theoretical guarantees
once small errors are involved. Thus while linear
regression uses an inappropriate error model for
samples from a power-law distribution, so too does
na\"ive application of maximum likelihood for samples
with routine and otherwise negligible error. In
principle, better MLEs could incorporate specifications
of the error distribution and its parameters
\citep{gillespie2017estimating}, but this requires
detailed knowledge of the specific dataset and a
perfect specification is impossible in practice.

\begin{figure}[t] \noindent
\includegraphics[width=89mm]{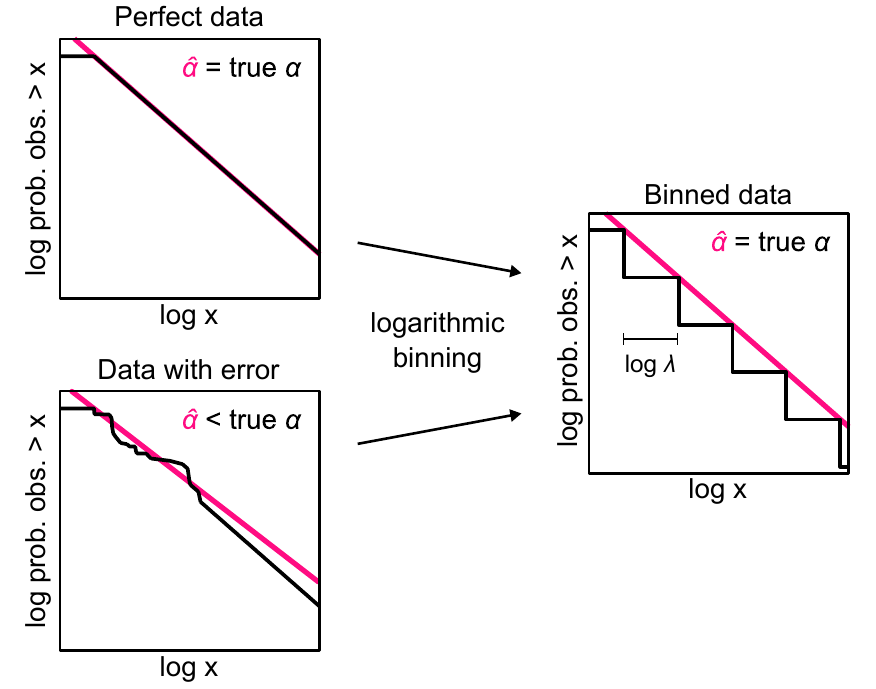}
\caption{Cartoon overview. 
Logarithmic binning causes power law
data with and without errors to converge toward the same
discrete power-law distribution.
Routine errors in data cause deviations from a
perfect power law that bias estimates
of the exponent $\hat\alpha$ and goodness-of-fit tests
(bottom left).  Logarithmic binning reduces the effect of
error by smoothing data within each bin. After binning,
data with errors better approximates perfect data from
the discrete power law.  Inference using the discrete
power law then recovers unbiased estimates and
stipulated false positive rates in goodness-of-fit
tests.}
\label{fig:cartoon}
\end{figure}

We offer a novel method to tune the robustness of MLEs
and K-S statistics to small errors without relying on
specific information about the errors. Rather than
model error explicitly, we reduce its affect by binning
the input data as depicted in Fig.~\ref{fig:cartoon}.
Small errors by definition typically preserve the order
of magnitude of each data point, and hence small errors
and relatively large bins rarely allow data to move
between bins, limiting the possible influence of
errors. The power law, meanwhile, specifies frequencies
across orders of magnitude, and so binning the data by
order of magnitude preserves much of the useful
information for inference.

Binning by orders of magnitude is a case of logarithmic
binning, where the bin boundaries are integer powers of
a ratio $\lambda > 1$. Logarithmic binning and the
power law distribution are both self-similar, and so
power law samples also follow the power law with the
original exponent after logarithmic binning
\citep{newberry2019self}. The discrete distribution of
the binned data has the same shape as the original, but
ignores errors that are small relative to the binning
ratio $\lambda$. Taking the limit of $\lambda \to 1^+$
recovers inference using the continuous Pareto
distribution. The customary Pareto inference method is
therefore an extreme case which is the most sensitive
to error. The discrete power law thus provides a more
robust general model for power-law distributed data,
recovering the benefits of MLEs and K-S statistics even
in the presence of small, unspecified errors.

We validate in simulation that logarithmic binning
attenuates biases in estimates as well as restores
specified false positive rates in goodness-of-fit tests
on power-law data with noise. Furthermore these
benefits can be achieved with a known and relatively
small cost in increased statistical error on the
estimator \cite{efron1978assessing,newberry2019self}.
We further find no impact on false negative rates in
rejecting non-power-law data for some binning schemes,
whereas others negotiate a tradeoff between false
positive and false negative rates. These results show
that logarithmic binning preserves most of the useful
information for parameter estimation and hypothesis
tests about power laws, while removing extraneous and
misleading effects of noise.

We further show that observable errors in empirical
datasets have caused biases in past inferences and
incorrect conclusions about whether data originates
from a power-law distribution. For example, errors such
as rounding to the nearest tenth, on either a linear or
logarithmic scale, can cause goodness-of-fit tests to
reject data that otherwise fits the power law, whereas
distributions with visually noticeable curvature across
their entire range can be accepted as a power law
unless binning induces the tests to adopt a broader
perspective. We therefore conclude with the
recommendation of logarithmic binning as a first step
of inference in power law distributions.

\section{Method}

We propose logarithmic binning as a smoothing method to
remove errors and reduce bias in parameter estimates
and measures of goodness of fit. Given the minimum
possible data value $x_m$, the logarithmic binning
scheme is fully specified with a continuous parameter
$\lambda > 1$ specifying the ratio between adjacent bin
boundaries. This bin width, $\lambda$, then controls
the amount of smoothing.  The limit $\lambda \to 1^+$
corresponds to no binning, since every unique data
point occupies its own bin and the binned and unbinned
data converge, whereas large $\lambda$ such as 2 or 10
bin the data by orders of magnitude and smooth out all
information within each order.

Given input data $x_i$, we assign the binned value
$\lfloor x_i\rfloor_\lambda = x_m\lambda^k$ where $k$
corresponds to the closest integer power of $\lambda$,
rounding down. We denote binning using a floor operator
with a subscript $\lambda$ because logarithmic binning
is a floor operation in $\log_\lambda$ space: in terms
of the usual integer floor, $\lfloor x \rfloor$,
logarithmic binning is $\lfloor x \rfloor_\lambda =
x_m\lambda^{\lfloor \log_\lambda x / x_m \rfloor}$.

We can derive an expression for the distribution of
binned data. The possible values after binning are
$x_m\lambda^k$ for $k = 0, 1, 2, ...$. By integrating
the probability density function of the continuous
power law (Eq.~\ref{eq:dcpl}) over the range of each
bin, the probability mass function for binned power-law
data is given by
\begin{equation}
p_d(x_m\lambda^k) =
\int_{x_m\lambda^k}^{x_m\lambda^{k+1}}
  \mkern-25mu p(x)\, dx = 
(1 - \lambda^{-\alpha})(\lambda^k)^{-\alpha}.
\label{eq:ddpl}
\end{equation}
This discrete distribution is also a power law
distribution since $\ln p_d = -\alpha \ln(x_m\lambda^k)
+ c$, with exponent $\alpha$ equal to the $\alpha$ in
Eq.~\ref{eq:tail}.  Logarithmic binning is the only
binning scheme that preserves this property
\citep{newberry2019self}. The continuous power law is
scale-invariant, whereas the discrete power law
(Eq.~\ref{eq:ddpl}) has a discrete-scale invariance
\citep{sornette1998discrete} with the same scaling
exponent.

For parameter estimation, we use the maximum likelihood
estimator for the discrete distribution given by
\citet{newberry2019self}, 
\begin{equation}
\hat\alpha_\lambda = \log_\lambda\left[1 +
\left({1 \over n}\sum_{i = 1}^n \left(\log_\lambda
\lfloor x_i \rfloor_\lambda-\log_\lambda x_m\right)\right)^{-1}\right],
\label{eq:mld}
\end{equation}
using the logarithmically-binned data $\lfloor x_i
\rfloor_\lambda$.  This estimator is notably undefined
for $\lambda = 1$.  However in the limit $\lambda \to
1^+$, $\hat\alpha_\lambda$ converges to the classical
MLE for the Pareto distribution due to
\citet{muniruzzaman1957measures}, and so we call this
estimator $\hat\alpha_1$,
\begin{equation}
\label{eq:mlc}
\hat\alpha_1 := 
\lim_{\lambda \to 1^+} \hat\alpha_\lambda = 
\left({1 \over n}\sum_{i = 1}^n
  \left(\log x_i-\log x_m\right)\right)^{-1}.
\end{equation}
Thus we index the estimator by the binning ratio
$\lambda \ge 1$, with $\lambda = 1$ representing the
case of raw, unbinned data.

The variance of the MLE $\hat\alpha_\lambda$ is given
by the inverse of the observed Fisher information
\citep{efron1978assessing,virkar2014power,newberry2019self}
as
\begin{equation}
\label{eq:sigmad}
\sigma^2_{\hat\alpha_\lambda} =
(\lambda^{\hat\alpha_\lambda} - 1)^2 /
(n\lambda^{\hat\alpha_\lambda} \log^2 \lambda). 
\end{equation}
This expression is also undefined for $\lambda = 1$,
but converges to $\sigma^2_{\hat\alpha_1} =
\hat\alpha^2_1/n$ in the limit $\lambda \to 1^+$ and
likewise corresponds to the variance on the Pareto MLE.

The variance on the estimator and hence the statistical
error increase polynomially with $\lambda$ as
$O(\lambda^{2\alpha})$. Conversely as $\lambda$
decreases, small errors are more likely to move data
between bins and bias the estimator.  Hence $\lambda$
negotiates a tradeoff between potential bias and
statistical error, with some intermediate optimum that
depends on the nature and severity of errors in the
data. The Pareto MLE $\hat\alpha_1$ in common use
occupies the extreme end of this spectrum with the
least variance but also the greatest susceptibility to
error.

As a goodness of fit measure, we use the
Kolmogorov-Smirnov statistic $D$. The
Kolmogorov-Smirnov statistic is the divergence between
a sample and a reference distribution computed as the
maximum difference between the empirical and null
hypothesized cumulative distribution functions. For
continuous distributions, the asymptotic distribution
of $D$, the Kolmogorov distribution, is classically
known exactly \citep{kolmogorov1933sulla}. The quantile
of the $D$ computed from a sample provides a measure of
goodness of fit to the distribution
\citep{massey1951kolmogorov}.  Extreme quantiles,
corresponding to low $p$-values, indicate discrepancies
between the sample and the hypothesized distribution.
The Kolmogorov-Smirnov test rejects the null
hypothesis, in favor of the alternative hypothesis that
the data originate from some other distribution, if the
$p$-value is less than the stipulated false positive
error rate. The $p$-value cutoff is equal to the
stipulated false positive rate because if a sample
truly originates from the reference distribution, the
distribution of its quantile and $p$-value are uniform.

The Kolmogorov-Smirnov statistic for
logarithmically-binned data $\lfloor x_i
\rfloor_\lambda$ relative to the discrete power law
(Eq.~\ref{eq:ddpl}) is
\begin{equation}
D =\, \max_{k \in \{0, 1, 2, ..., \infty\}} \left|{1
\over n}\sum_{i=1}^n \mathbf{1}_{\lfloor x_i \rfloor_\lambda \le
x_m\lambda^k} - \sum_{i=0}^k p_d(x_m\lambda^k)\right|,
\end{equation}
where the sum over the indicator function $\mathbf{1}$
counts the number of data points in bins up to and
including $x_m\lambda^k$. The resulting distribution of
$D$, however, is not necessarily equal to the
Kolmogorov distribution when the null hypothesized
distribution is discrete or involves parameters
estimated from the data
\citep{noether1963note,walsh1963bounded,lilliefors1967kolmogorov}.
Therefore we build up an approximate distribution of
$D$ under the null hypothesis by bootstrapping
following \cite{lilliefors1967kolmogorov}. We draw
a sample of size $n$ from the discrete power law given
by Eq.~\ref{eq:ddpl} with parameters $\lambda$ and
$\hat\alpha_\lambda$, fit the parameter
$\hat\alpha_\lambda'$ to this sample, then compute $D$
for this sample using the parameters $\lambda$ and
$\hat\alpha_\lambda'$, thus simulating the steps for
computing $D$ from the data. This procedure can be
repeated on many samples from the discrete power law to
build up an empirical distribution of $D$ that
successively better approximations to the true null
distribution. The $p$-value is one minus the quantile
of $D$ computed from the data among the samples of $D$
computed through bootstrapping.  This goodness-of-fit
test follows a popular method
\citep{clauset2009power,gillespie2015fitting} that has
also been adapted to binned power-law data
\citep{virkar2014power}. Our test differs only in that
we do not jointly estimate $x_m$, as our analysis fixes
$x_m = 1$.  In lieu of computing precise $p$-values by
bootstrapping many samples, we just as accurately judge
whether $p < 0.05$ using only 19 bootstrapped values,
concluding that $p < 0.05$ if the $D$ observed in the
data exceeds all 19.

This logarithmic binning method allows the experimenter
to specify $\lambda$ based on tolerances of bias and
statistical error. The magnitude of bias cannot be
estimated directly for unknown sources of error. We
provide tolerances (see Validation) for the case of
additive and multiplicative normal noise. In principle,
bias can be minimized by choosing $\lambda$ to exceed
relative error on the vast majority of data points or
to choose $\lambda$ as high as possible. Upper limits
for $\lambda$ depend on the application. A given
tolerance for statistical error imposes an upper limit
on $\lambda$ given by Eq.~\ref{eq:sigmad}. Even with
unlimited tolerance for statistical error, data must
occupy at least two bins for $\hat\alpha_\lambda$ to be
defined. The number of bins is equal to $\lfloor
\log_\lambda \max_i x_i / x_m \rfloor + 1$.  Hence the
maximum feasible $\lambda$ is equal to the proportional
range of the data, $r := \max_i x_i / x_m$.  The
corresponding $\hat\alpha_r$ minimizes bias but also
has the highest possible statistical error and is
useless for foreseeable practical applications.
Hypothesis testing imposes more restrictive upper
limits on $\lambda$. Generally, $\lambda$ should be
constrained by the capacity to reject alternative
distributions. What values this constraint imposes
unfortunately depends on the alternative distribution,
which is often unspecified. One extreme upper limit is
given by requiring at least three bins. With data in
only two bins, $\hat\alpha_\lambda$ can be typically be
chosen to fit the data exactly so that $D < 1/n$ in
every bootstrap sample. The probability a bootstrap run
contains fewer than three bins is given by the formula
\begin{equation}
\Pr(\max_i x_i < \lambda^2) =
\left(1 - \left(\lambda^2\over x_m\right)^{-\alpha}\right)^n
\label{eq:ullambda}
\end{equation}
derived from Eq.~\ref{eq:tail}.  Setting this
expression equal to the tolerable fraction of bootstrap
simulations with fewer than three bins gives the
theoretical upper limit on $\lambda$. In practice,
other considerations such as false negatives rates will
further restrict the upper limit of $\lambda$ for
purposes of hypothesis testing.

\section{Validation}

We validate the performance of binning by synthesizing
pure continuous power-law samples and introducing
normal additive and multiplicative (proportional)
noise. We generate parameter estimates and goodness of
fit measures for different levels of binning from no
binning to the extremes allowed by the data. We compare
the known, true parameters to the estimated parameters
and compare the sensitivity and specificity of
goodness-of-fit tests with and without binning.

\begin{figure}[t] \noindent
\hspace{-20mm}\includegraphics[width=180mm]{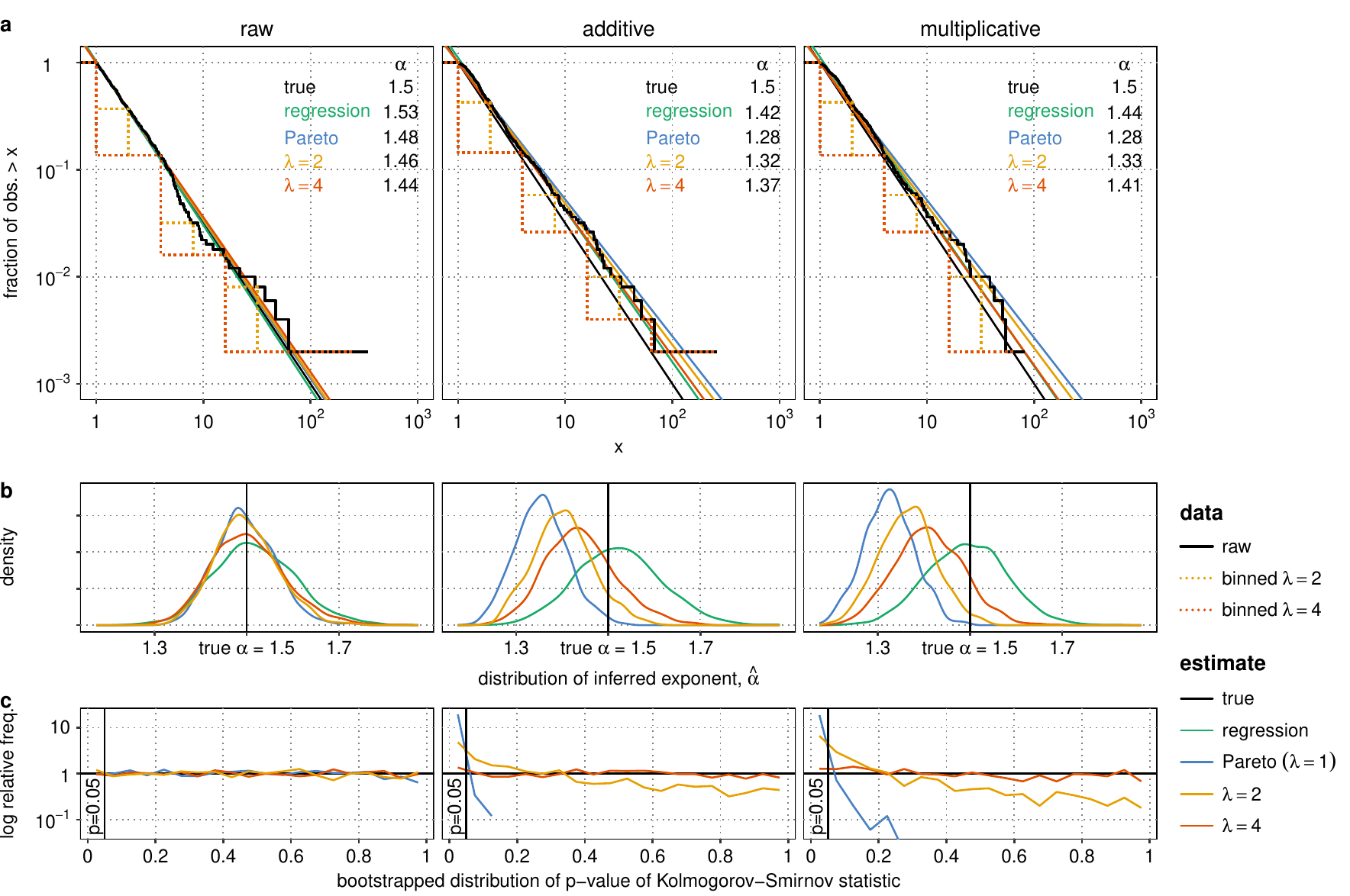}
\caption{Error in power-law data biases estimates of
the exponent $\alpha$ and causes tests based on
Kolmogorov-Smirnov (K-S) statistics to reject the
power-law distribution.  (a) Log-log tail distribution
plots of samples from perfect data (left) and data with
additive and multiplicative noise are visually almost
indistinguishable (middle: Normal(0, $\sigma_+^2$=0.2),
right: Lognormal(0, $\sigma_\times^2$=0.2)).  This
noise nonetheless biases estimates of the log-log slope
$-\alpha$ relative to the true $\alpha$=1.5.
Logarithmic binning with $\lambda$=2 or $\lambda$=4
(dotted lines) brings the slope estimates
$\hat\alpha_\lambda$ closer to the true value. (b)
Distributions of the slope estimates over 1000 samples
of size $n=500$ illustrate a tradeoff between accuracy
and precision, where the most precise estimation
methods are also the most inaccurate. (c) Distributions
of $p$-values for rejecting the power law based on a
K-S statistic are uniform when the data comes from a
perfect power law, with or without binning (left).
However, noise biases $p$-values (middle, right) so
that the K-S statistic that assumes a continuous power
law (blue) has $p$<0.05 more than 50\% of the time.
Binning with $\lambda$=2 or $\lambda$=4 attenuates
noise, brings $p$-value distributions closer to
uniformity, and restores stipulated false positive
rates.}
\label{fig:fits}
\end{figure}

We synthesize data by drawing $n$ independent samples
$x_i$ from a Pareto distribution (Eqs.~\ref{eq:tail},
\ref{eq:dcpl}) with parameters $\alpha$ and $x_m$.
Without loss of generality, we set $x_m$ to 1, since
the general case can always be mapped to $x_m = 1$ by
converting the units of $x$. We simulate experimental
error using either unbiased additive (normal) or
multiplicative (lognormal) noise with variance
$\sigma_+^2$ and $\sigma_\times^2$ respectively. The
data with additive noise is constructed from a sample
$x_i$ as $x_i + \operatorname{Normal}(0, \sigma_+)$ and
multiplicative noise as
$x_i\times\operatorname{Lognormal}(0, \sigma_\times)$.
For $x_m \ne 1$, the corresponding additive variance is
$x_m\sigma_+$, since $\sigma_+$ has the same units as
$x$, whereas the corresponding multiplicative variance
is still $\sigma_\times$ since $\sigma_\times$ is a
dimensionless log-ratio.  Fig.~\ref{fig:fits}a shows
tail distribution plots of samples generated with each
kind of noise.

As a proof of principle, we take 1000 samples of size
$n = 500$ and $\alpha = 1.5$ with and without additive
and multiplicative noise with $\sigma_+ = \sigma_\times
= 0.2$ in order to generate empirical distributions of
estimates $\hat\alpha$ and bootstrapped $p$-values of
the K-S statistic (Fig.~\ref{fig:fits}).  For each
sample we estimate $\alpha$ and compute the $p$-value
both without binning (Pareto) and using logarithmic
binning with $\lambda = 2$ and $\lambda = 4$. We
additionally estimate $\alpha$ as the slope by ordinary
least squares regression on the empirical cumulative
distribution function---the log of each data point
versus the log of its quantile in the sample, for which
no binning is necessary.

Fig.~\ref{fig:fits} shows how much the errors can bias
estimates of $\alpha$ and $p$-values. All estimates and
$p$-value behave as expected in samples without error:
the estimates cluster around the true value and
distributions of $p$-values are uniform regardless of
$\lambda$.  The averages of each MLE
$\hat\alpha_\lambda$ are within 0.21\% of the true
value whereas the average regression estimate is within
1.2\%. In samples with noise however, the na\"ive
Pareto MLE $\hat\alpha_1$ is biased by around 10\%,
confidence intervals on $\hat\alpha_1$ typically
exclude the true value, and $p$-values approach zero.
This observation parallels empirical findings that
errors in the data such as measurement error or
quantization noise can substantially bias MLEs in
practice \citep{langlois2014maximum,newberry2019self}.

Coarser binning, with larger values of $\lambda$,
brings the MLEs $\hat\alpha_\lambda$ closer to the true
value on average and brings the distribution of
$p$-values closer to uniform. In all noise treatments,
binning with $\lambda = 4$ yields approximately uniform
$p$-values and regression provides approximately
unbiased estimates of $\alpha$.

Bias and variability of estimates of $\alpha$
illustrate a clear tradeoff between accuracy and
precision. As expected for maximum likelihood
estimation, $\hat\alpha_1$ is the most efficient and
the distribution of $\hat\alpha_1$ is the most sharply
peaked (Fig.~\ref{fig:fits}b). However, $\hat\alpha_1$
is correspondingly the most sensitive to errors in the
data and thereby produces the most biased estimates.
Binning makes the $\hat\alpha_\lambda$ progressively
less sensitive to error for $\lambda = 2$ and $\lambda
= 4$, while providing more precise estimates than
linear regression. Estimates by linear regression are
the most accurate as well as the most variable. On the
whole, linear regression minimizes the squared
difference between the estimates and the true value in
this example, despite its lack of theoretical
guarantees.

Noise also causes the bootstrapped K-S $p$-values to
reject the Pareto distribution nearly all of the time
(additive: 97\% multiplicative: 93\%), in contrast to a
stipulated false positive rate of 5\%
(Fig.~\ref{fig:fits}c). In one sense, the test is
correctly performing its job: the null hypothesis
specifies that the sample originates from the Pareto
distribution, whereas the distribution of the sample is
Pareto convolved with noise. By a strict
interpretation, the null hypothesis is correctly
rejected. In a more practical sense however, the data
comes from a power law whether or not small errors are
invovled, and so rejecting the power law is incorrect
and constitutes a false positive.  All measurements
have errors, and so in this sense, which we adopt here,
errors in the data obscure the truth by causing extreme
bias in the $p$-values.

Binning the data brings $p$-values back to an
approximately uniform distribution, as shown in
Fig.~\ref{fig:fits}c. Binning blinds the test to
differences in the shape of the empirical and
hypothesized distributions within each
scale---including differences due to discretization and
measurement errors---while still accounting for the
shape of the distribution across the range of the data.
With $\lambda = 2$, the false positive rates fall to
23\% and 34\% on data with additive and multiplicative
noise, respectively. With $\lambda = 4$, the rejection
rates in these 1000 trials are consistent with the
stipulated rejection rate of 5\% at 5.4\% and 6.1\%
respectively.  More trials would reveal biases on the
order of 1\%, but these K-S $p$-values are valid for
purposes of roughly controlling the false positives
rate. That is, after binning by sufficiently large
$\lambda$, data originating from a power law, errors or
not, will be rejected as a power law with $p<0.05$
roughly 5\% of the time.

\begin{figure}[t] \noindent
\centering\includegraphics[width=90mm]{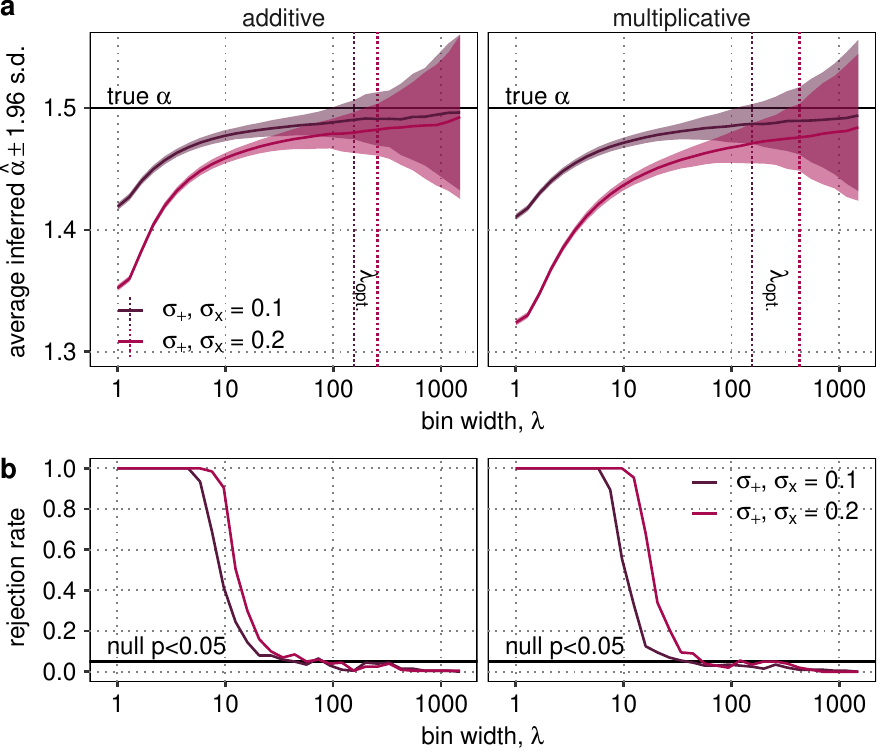}
\caption{The binning ratio $\lambda$ tunes (a) a
tradeoff between accuracy and precision of the
estimator $\hat\alpha_\lambda$ and (b) robustness of
hypothesis tests, on data with additive and
multiplicative noise.  Here we simulate large datasets
($n$=1,000,000, median range $r$=14,267, 200
replicates) with additive and multiplicative noise with
variance $\sigma_+$ and $\sigma_\times$ = 0.1 and 0.2.
(a) Increasing $\lambda$ brings estimates
$\hat\alpha_\lambda$ closer to the true $\alpha$=1.5
but also increases the variance of the estimator. The
optimal binning ratio $\lambda_{\mathrm{opt.}}$ to
minimize total mean squared error on the estimate
depends on the type and magnitude of noise but roughly
divides the data into two bins.  (b) Hypothesis tests
virtually always reject the power law unless $\lambda$
is sufficiently large relative to noise. Rejection
rates roughly equal the stipulated $p$-value cutoff
0.05 when $\lambda \approx 70$ in all noise treatments
whereas yet greater values of $\lambda$ produce
conservatively biased $p$-values. The range of
$\lambda$ that achieves the stipulated false positive
rate corresponds to dividing the data into 3-4 bins.}
\label{fig:lambda}
\end{figure}

The binning ratio $\lambda$ controls the tradeoff
between accuracy and precision in estimates and the
robustness of hypothesis tests to noise, as shown in
Fig.~\ref{fig:lambda}. Increasing $\lambda$ attenuates
noise but also removes information from the sample.
Biases in $\hat\alpha_\lambda$ therefore decrease with
$\lambda$ whereas variability of the estimator
(Eq.~\ref{eq:sigmad}) increases.  We empirically
investigated this tradeoff on large samples by fitting
$\hat\alpha_\lambda$ and conducting hypothesis tests
across a range of $\lambda$ (200 replicates;
$n$=1,000,000; $\alpha = 1.5$; $\sigma_+, \sigma_\times
= 0.1, 0.2$).  For each noise treatement we found a
$\lambda_{\mathrm{opt.}}$ that minimized the total mean
squared error on the estimator, $(\alpha -
\hat\alpha_\lambda)^2$, incorporating both bias and
variability.  This $\lambda_{\mathrm{opt.}}$ typically
divided the data into two bins. Specifically, setting
$\log \lambda / \log r = 0.50$ with $r = \max_i x_i /
x_m$ equally divides the range of the data in log space
(median: $r$ = 14,267, $\lambda_{0.50} = 119$). We
observed $\log \lambda_{\mathrm{opt.}} / \log r$
ranging from 0.49 to 0.63 across all noise treatments.
Doubling the noise leads to only marginally larger
$\log \lambda_{\mathrm{opt.}} / \log r$, suggesting
that equally dividing the range of the data into two
bins is a reasonable generic prescription for
minimizing bias and variability when detailed
information about errors in the data is not available.

Binning with sufficiently large $\lambda$ restored
stipulated false positive rates in hypothesis tests
(Fig.~\ref{fig:lambda}b). On these large samples, the
hypothesis test nearly always rejects the power law
unless $\lambda$ is also large. The minimum $\lambda$
to achieve the stipulated false positive rates depends
on the type and magnitude of noise and ranged from
$\lambda = 20$ to $\lambda = 60$. For yet higher
$\lambda$, $p$-values become conservatively biased from
too few bins (see Methods), causing fewer than the
stipulated rate of false positives. In all four
treatments, $\lambda$ in the range from 60 to 75 gave
rejection rates that were not significantly biased
upwards or downwards from the stipulated rate of 0.05
over 200 trials (one-tailed binomial $p$-values: 0.12
to 0.87). According to Eq.~\ref{eq:ullambda}, the bias
from too few bins at $\lambda = 75$ is less than 10\%
or 0.005 while the bias due to errors in the data is
minimized. Thus robust $p$-values are available for
$\lambda = 75$ across all treatments as well as for
lower $\lambda$ in treatments with lower error.

\begin{figure}[t]
\noindent
\centering
\includegraphics[width=90mm]{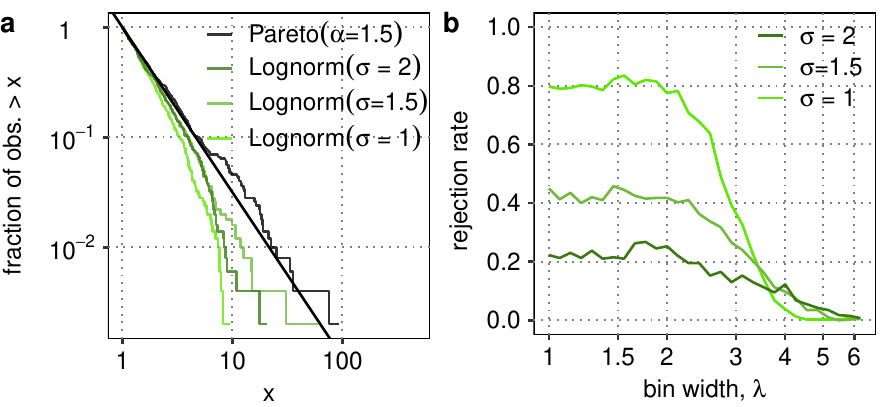}
\caption[]{For (a) lognormal samples chosen to resemble
the power-law, (b) binning entails no decrease in
statistical power below a threshold of $\lambda$.  (a)
Samples ($n$=500) from the tail of the lognormal
distribution can approximate power-law samples
arbitrarily well by increasing the lognormal variance
$\sigma$. (b) The statistical power (rejection rate) on
these samples is unaffected by binning provided
$\lambda < 2$, corresponding to dividing the data into
at least four bins.}
\label{fig:falseneg}
\end{figure}

Statistical power or sensitivity is the test's ability
to correctly reject non-power-law data. Intuitively
sensitivity might decrease with $\lambda$ since binning
removes information. We investigate how binning affects
the ability to reject non-power-law data by applying
the test to small samples ($n = 500$) from the tail of
the lognormal distribution truncated below $x_m = 1$.
The lognormal distribution, like the power law, is
heavy-tailed and so its extremes can be difficult to
distinguish from a power law. Both tail distributions
can appear straight on a log-log plot, but the
lognormal has curvature that depends on its parameters
(Fig.~\ref{fig:falseneg}a). The truncated lognormal
distribution has three parameters, the mean and
variance $\mu$ and $\sigma$ as well as the tail
threshold $x_m$ that samples must exceed.  The slope
and curvature of the lognormal tail depend on all three
parameters. We set $x_m$ to 1, use $\sigma$ to set the
curvature, and finally choose $\mu$ so that the log-log
slope of the lognormal tail distribution at $x_m$ is
equal to a power law with $\alpha = 1.5$. In this
scheme, the lognormal tail can approximate the power
law arbitrarily well as the lognormal variance $\sigma$
increases.

For samples from the lognormal tail, we find that the
test sensitivity is unaffected by binning, provided
$\lambda$ divides the data into at least four bins
(Fig.~\ref{fig:falseneg}b). The test's rejection rate
on lognormal data depends strongly on the curvature of
the lognormal distribution and the amount of data. For
$\sigma = 1$, the test rejects lognormal data roughly
80\% of the time regardless of $\lambda$ for all
$\lambda < 2$. The median range of these samples is
12.8, and so $\lambda = 2$ typically puts the maximum
data point in the fourth bin, [8,16). As $\sigma$
increases, so too does the maximum value in samples,
which widens the range of $\lambda$ that divide the
data into four bins.  One quick, intuitive
rationalization for the significance of four bins is
that only three points are required to detect
curvature, while the topmost bin is unreliable and only
partially occupied by the range of the data.

Chosing $\lambda$ within a certain range tunes a
tradeoff between sensitivity and specificity. Our
earlier measurements in noisy power-law samples of the
same size ($n$=500, Fig.~\ref{fig:fits}) give the
specificity. While the sensitivity is the rate of
rejection on lognormal tail samples, the specificity is
the rate of acceptance on noisy power-law samples or
one minus the false positive rate.  The test with
$\lambda = 2$ and a $p$-value cutoff of 0.05
distinguishes lognormal tail samples with $\sigma = 1$
from noisy power-law samples with a sensitivity of 78\%
and specificities of 77\% for additive noise and 66\%
for multiplicative with $\sigma_+, \sigma_\times =
0.2$.  Using $\lambda < 2$ yields reduced specificity
with no benefit to sensitivity, whereas $\lambda > 2$
tunes a tradeoff between sensitivity and specificity.
Using $\lambda = 3$ for example gives sensitivity of
37\% and specificities of 92\% and 89\% on the same
samples. For $\lambda > 4$ sensitivity approaches zero
and the test is useless.  The interval $2 < \lambda <
4$ then offers reasonable tradeoffs between sensitivity
and specificity, with lower $\lambda$ more sensitive
and higher $\lambda$ more specific.

\begin{figure}[t]
\noindent
\hspace{-20mm}\includegraphics[width=180mm]{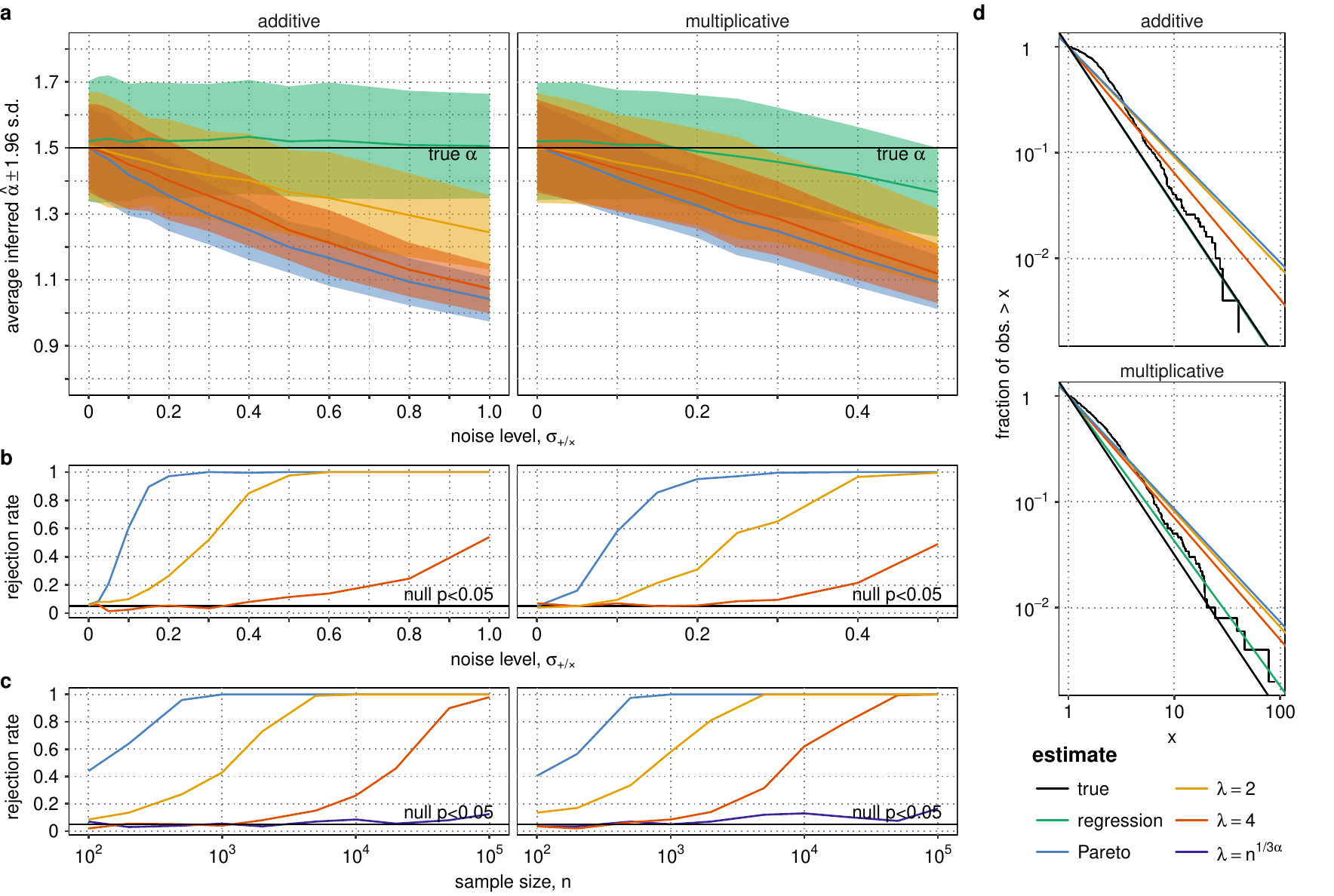}
\caption{Increasing noise $\sigma_+$ and
$\sigma_\times$ (a) increases bias in estimates of
$\alpha$ as well as (b) the rate of rejecting the
power-law distribution in samples of size $n=500$,
while (c) increasing sample size $n$ increases the
rejection rate without affecting bias in
$\hat\alpha_\lambda$ in samples with $\sigma_+,
\sigma_\times = 0.2$. At the upper limit of the plotted
range, where $\sigma_+ = 1.0$ and $\sigma_\times =
0.5$, (d) the tail distributions of samples are notably
curved and no longer resemble a power law on log-log
plots, producing extremely biased estimates of the
slope. Binning with $\lambda = 2$ or 4 (yellow, red)
reduces bias in estimates and restores rejection rates
to the stipulated 0.05 over a larger range of tolerable
error.}
\label{fig:sigman}
\end{figure}

The foregoing results demonstrate cases in which
binning has the desired effect of reducing the
influence of small errors in the data at modest cost of
statistical power. The qualitative results hold more
generally, while exact magnitudes of bias and rejection
rates depend on the errors, the sample size, and
$\alpha$. The test sensitivity and how it depends on
$\lambda$, on the other hand, depend on the possible
alternative distributions, which we cannot enumerate
explicitly.

We can quantitatively generalize the results by varying
$\sigma$ and sample size $n$. Biases in $\alpha$
increase with greater error magnitudes and decrease
with greater $\lambda$, while false positive rates as
well as the feasible values of $\lambda$ depend
strongly on sample size (Fig.~\ref{fig:sigman}).  We
choose a range of errors from 0 (no error) to a maximum
error high enough that the distribution clearly
deviates from a straight line on a log-log plot and no
one slope clearly represents the distribution
(Fig.~\ref{fig:sigman}b). The error rates $\sigma_+$
and $\sigma_\times$ each have marginal effects on
biases in $\alpha$ (Fig.~\ref{fig:sigman}a), whereas
the biases do not depend on $n$ because greater sample
size merely causes the estimators to converge to their
expected values. Precision of the estimates, by
contrast, increases with sample size, as expected from
Eq.~\ref{eq:sigmad}. This feature can be problematic
because as sample size increases, estimators are more
confident in an incorrect estimate and confidence
intervals are more likely to exclude the true value.
Fortunately increasing $n$ also increases the feasible
range of $\lambda$ and the maximum $\lambda$ subject to
a tolerable statistical error.  Hence increasing sample
size can mitigate bias by concommittantly increasing
$\lambda$ according to Eq.~\ref{eq:sigmad}.

Rejection rates exhibit sigmoid threshold behavior in
both $\sigma$ and $n$ similar but opposite to the
thresholds in $\lambda$ (Fig.~\ref{fig:sigman}c,
Fig.~\ref{fig:lambda}). For sufficiently small error or
sample size, the test cannot detect errors, $p$-value
distributions are close uniform, and hence rejection
rates are equal to the $p$-value cutoff such as 0.05.
As error or sample size increases, hypothesis tests are
more likely to detect the errors and reject the power
law until eventually the errors or sample size are
sufficient for the test to nearly always reject.
Increasing $\lambda$ increases the threshold of
$\sigma$ or $n$ at which the test begins to detect
errors and reject the power law.  With $\lambda = 1$
(no binning) there is almost no range of tolerable
error.  For example, noise with $\sigma_+ = 0.025$
roughly doubles the false positive rate. For $\lambda =
4$, on the other hand the noise must exceed $\sigma_+
\ge 0.4$ or $\sigma_\times \ge 0.3$ before the
rejection rate is doubled. While noise can easily be
increased to the point that no feasible binning is
sufficient to restore the stipulated rejection rates,
increasing $n$ allows greater $\lambda$ by increasing
the range of the data. In Fig.~\ref{fig:sigman}, we set
$\lambda$ to $n^{1/(3\alpha)}$ (purple), designed to
typically bin the data into 3 or 4 bins.  Allowing this
$\lambda$ to depend on $n$ preserves approximately
unbiased $p$-values over the range of $n$ from 100 to
100,000.

\begin{figure}[t]
\noindent
\hspace{-20mm}\includegraphics[width=180mm]{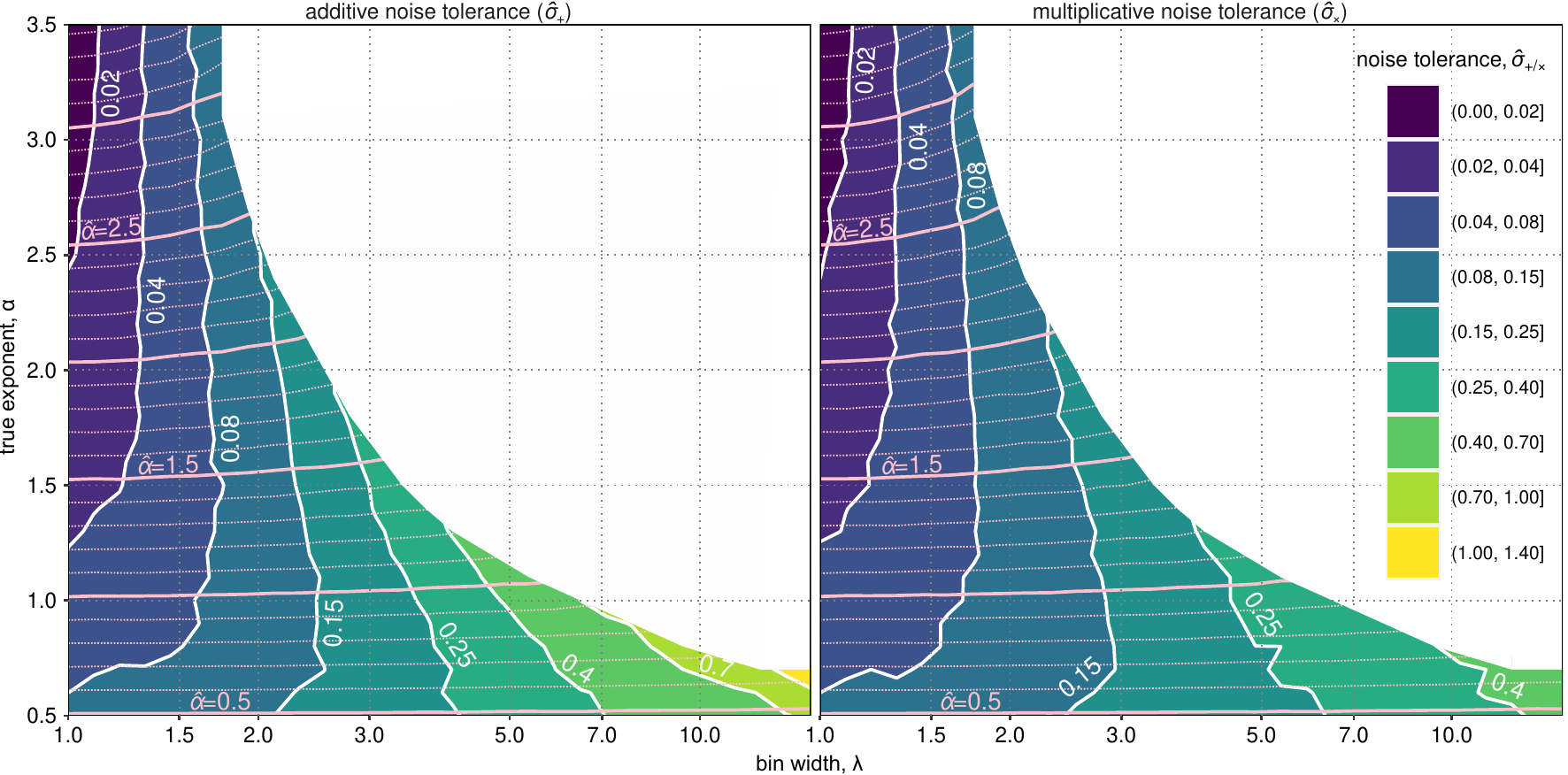}
\caption[]{The noise tolerance for a given $\alpha$ and
$\lambda$ is the level of noise $\sigma_+$ or
$\sigma_\times$ at which bootstrapped K-S $p$-values
are less than 0.05 roughly 10\% of the time. Shown here
are noise tolerance contours for samples of size
$n$=500. With this amount of error, $p$-values are
biased enough to double the false positive rate
relative to the stipulated rate of 0.05. More error
quickly renders hypothesis tests meaningless, while
less error, greater $\lambda$, or lower $n$ can restore
the actual false positive rates closer to the
stipulated 5\%. The inferred $\hat\alpha_\lambda$ (pink
contour lines) are also biased to some degree by this
amount of noise, as they differ from the horizontal
lines of true $\alpha$.  Bias in $\hat\alpha_\lambda$
increases with $\sigma_{+/\times}$ but decreases with
$\lambda$ so that as $\sigma_{+/\times}$ increases with
$\lambda$, the contour lines remain roughly horizontal
and bias remains within 10\% of the true $\alpha$.}
\label{fig:alpha}
\end{figure}

The parameter $\alpha$ also influences the error
threshold and magnitudes of bias in the MLEs.  We
varied $\alpha$ and $\lambda$ over a grid of
combinations and computed error thresholds as the
$\sigma_+$ or $\sigma_\times$ required to double the
rejection rate for a $p$-value cutoff of 0.05. For each
$\alpha$ we varied $\lambda$ from 1 to $n^{1/3\alpha}$,
which at its maximum still divided the data into at
least four bins at least 70\% of the time.  We
conducted a stochastic binary search to estimate these
values $\hat\sigma_+$ and $\hat\sigma_\times$, testing
various candidate $\sigma$-values and concluding the
search when the target rejection rate 0.10 was within a
binomial 95\% confidence interval of $\pm 0.005$ for
trials in the interval $\hat\sigma\pm0.02\hat\sigma$.
This search thus yields values $\hat\sigma$ which are
within 2\% of some $\sigma$ for which the rejection
rate is within 5\% of 0.10 approximately 95\% of the
time. The search also yields mean estimates
$\hat\alpha_\lambda$ at the inferred noise level.

Fig.~\ref{fig:alpha} thereby gives an error tolerance
for hypothesis tests and the maximum bias on estimates
within that error tolerance applicable to samples of
$n=500$.  For example, at the values $\alpha = 1.5$ and
$\lambda = 2$, additive noise with $\sigma_+ < 0.08$
gives a rejection rate less than 0.1 and
$\hat\alpha_\lambda$ slightly less than 1.5, biased by
approximately 0.05 corresponding to results presented
in Fig.~\ref{fig:sigman}.  The lines of inferred
$\hat\alpha_\lambda$ are approximately horizontal and
approximately equal to $\alpha$ because limiting noise
to levels that do not substantially affect hypothesis
tests also limits noise that would substantially bias
$\hat\alpha$. At the maximum such $\hat\sigma_+$ or
$\hat\sigma_\times$ for any given $\alpha$, the
magnitude of bias $\hat\alpha - \alpha$ is still less
than 10\% of $\alpha$ when binning by the corresponding
$\lambda$.

Conversely, for a known amount of noise,
Fig.~\ref{fig:alpha} gives a lower bound for $\lambda$
at a given inferred $\alpha$. For example, on data with
a known $\sim$10\% proportional random noise,
$\sigma_\times \approx 0.1$ and hence, if inferred
$\alpha = 1.5$, $\lambda$ should be chosen to exceed
roughly 2.2 in order to remain within a 10\% tolerance
of bias on $\alpha$ and a less-than-doubled false
positives rate. Normal additive and multiplicative
noise can be taken as a proxy for many kinds of
measurement errors that routinely occur in data, and so
Fig.~\ref{fig:alpha} provides lower bounds on valid
$\lambda$ for a variety of error types in samples of
size $n=500$.

\section{Results}

We analyse three empirical cases and compare results
with and without binning. We use data on earthquake
magnitudes and wealth, which are historically believed
to follow power-law distributions, as well as wildfire
size, for which we have no evidence of a power law. The
specific datasets are curated online by
\cite{clauset2009power} as demonstration cases of
power-law inference.  The earthquakes dataset contains
17,450 positive and valid samples, recorded as
two-digit Richter magnitudes ranging from 0.5 to 7.8.
The Richter scale records the logarithm to the base 10
of the amplitude of waves recorded by seismographs
\citep{richter1935instrumental}.  We convert the data
back to the natural scale by exponentiating the Richter
magnitudes. The result is a dataset with discrete
values proportional to integer powers of
$\lambda=10^{0.1} \approx 1.26$. In other words,
Richter magnitudes with two digits of precision are
inherently a case of logarithmic binning with
$\lambda=10^{0.1}$.  The wealth dataset is the Forbes
list of the world's 399 richest people in 2003,
including 261 billionaires. The wildfires dataset
includes 203,784 measurements of wildfire area from one
decade in the US, 99\% of which are between $10^{-1}$
and $10^3$ acres.

\begin{figure}[t] \noindent
\hspace{-20mm}\includegraphics[width=180mm]{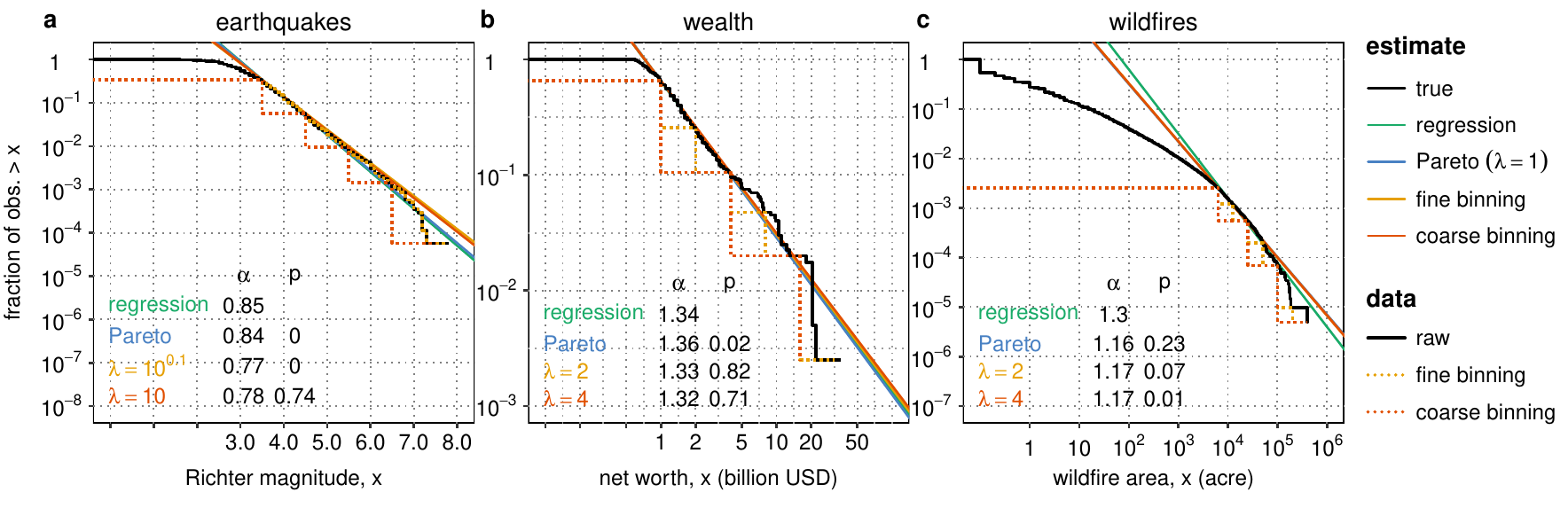}
\caption{Binning reverses conclusions about whether (a)
earthquake magnitudes, (b) wealth, and (c) wildfire
area follow a power law distribution.  Measurement or
recording errors and quantization noise (a,b) bias MLEs
and invalidate goodness-of-fit tests whereas binning
the data attenuates the influence of errors and induces
more reasonable conclusions.  For data that does not
follow a power law (c), binning correctly rejects the
power law when the test would otherwise accept it.}
\label{fig:empirical}
\end{figure}

We estimate the exponent $\alpha$ and conduct
goodness-of-fit tests for all three datasets using the
MLEs $\hat\alpha_1$ and $\hat\alpha_\lambda$ and K-S
statistics just as with the synthetic data. For each
dataset we compare three binning treatments---no
binning ($\lambda = 1$), fine binning (small $\lambda$)
and coarse binning (large $\lambda$)---letting the
small and large $\lambda$ depend on the dataset. We
also estimate the exponent with linear regression
against the empirical cumulative distribution function
as with the synthetic data in Fig.~\ref{fig:fits}.  We
stipulate a fixed $x_m$ for each dataset, specifically,
magnitude 3.5 for earthquakes, one billion dollars for
wealth, and 6324 acres for wildfires.

The earthquake dataset incorporates errors known to
complicate measurements, including partial censorship
of small values, quantization error, and attraction to
particular values. Fig.~\ref{fig:empirical}a
demonstrates that these subtle errors have different
effects on MLEs and goodness-of-fit tests at different
$\lambda$.  We avoid the partial censorship that causes
visible curvature in Fig.~\ref{fig:empirical}a by
choosing $x_m = 10^{3.5}$, or magnitude 3.5, across all
treatments. This minimum magnitude without censorship
is called the magnitude of completeness or $m_c$ in
seismology \citep{woessner2005assessing} and depends on
the density of the earthquake sensor network.  The
earthquake dataset also includes quantization error
from the two-digit precision of Richter magnitudes.
This quantization error is equivalent to logarithmic
binning, and so we choose the fine bin width as
$\lambda = 10^{0.1}$ to exactly match the quantization
in the raw data. This ``binning'' preserves the input
data exactly but entails a different estimator,
$\hat\alpha_{10^{0.1}}$, and a different bootstrapped
distribution of K-S statistics. Hence the fine-binned
($\lambda = 10^{0.1}$) and unbinned ($\lambda = 1$)
empirical cumulative distributions in
Fig.~\ref{fig:empirical}a overlap perfectly.  The
effect of quantization error \textit{per se} is then
evident from the difference between the Pareto MLE
$\hat\alpha_1 = 0.84$, which assumes no quantization,
and the appropriate MLE for the raw data,
$\hat\alpha_{10^{0.1}} = 0.77$. These values differ by
0.07 or 9\%, comparable to the 12\% magnitude of bias
due to quantization error in a different earthquake
dataset \citep{newberry2019self}.

We also observe abnormalities in the earthquake dataset
at special magnitudes 3.0, 3.5, 4.0, 4.5, etc., each of
which contains markedly more observations than adjacent
magnitudes. The effect is highly statistically
significant ($p<0.0001$, 1 d.f. $\chi^2$ test for
association between multiples of 0.5 and
higher-than-previous values) and easily explained if
the dataset combines low-precision data quantized by
0.5 with high-precision data quantized to 0.1. Binning
using coarse bins with $\lambda = 10$ corresponding to
a difference of 1.0 in Richter magnitude completely
removes this artifact. Coarse and fine binning provide
consistent estimates
$\hat\alpha_{10^{0.1}}\approx\hat\alpha_{10}=0.78$ even
with wildly different bin widths, but come to opposite
conclusions in goodness-of-fit tests: $p$=0.76 with
$\lambda=10$ whereas otherwise $p$<0.01. Indeed, with
no binning or fine binning by $\lambda=10^{0.1}$, the
artifact persists. This data is inconsistent with a
power law simply because more observations occur at
increments of 0.5 than could be explained by chance.
With $\lambda=10$, however, the artifact is removed and
we conclude that the earthquake data is consistent with
a power law. This binning still divides $n$=5,910
observations into 5 bins---offering plenty of capacity
to reject alternative distributions relative to our
simulations in which $n$=500 and 4 bins were sufficient
to reject lognormal data. We therefore conclude that
despite its flaws, the data provide strong evidence
that the underlying phenomenon of earthquake magnitudes
is consistent with power-law scaling across earthquake
magnitudes 3.5 to 7.8, in agreement with geological
studies and conventional wisdom.

The wealth dataset contains artifacts due to
discretization on a linear scale, as well as known and
unknown limitations on accurate measurements of the
wealth of the extremely rich
\citep{piketty2013capital}. We choose $x_m =
10^9$---one billion US dollars---for the Forbes list's
celebrated reputation for tracking the wealth of
billionaires.  This dataset includes striking
discrepancies in its quantification scheme, with values
truncated to the nearest 5 million or 100 million
depending on whether net worth exceeded one billion,
evident in different levels of jaggedness below and
above 1 in the curve of Fig.~\ref{fig:empirical}b.  The
log-log plot also suggests inconsistent representation
of values between 4 billion and 10 billion, where many
billionares are represented has having either
``4,000,000'' or ``5,000,000'' while others are
distinguished between 9.7 and 9.8 billion, particularly
if net worth is just shy of a round number, resulting
in noticeable ``bumps'' in the tail distribution.  In
this dataset, all three MLEs and regression obtain
similar $\alpha$ estimates, but goodness-of-fit tests
with and without binning draw distinct conclusions.  As
we observe in the earthquakes data, quantization noise
can be sufficient for K-S $p$-values bootstrapped from
the continuous Pareto distribution to reject the power
law.  Without binning, the MLE $\hat\alpha_1$ produces
a slope aligned with a majority of data points in the
dataset and consistent with the other estimates, but
the K-S statistic rejects the hypothesis that wealth is
drawn from a power-law distribution with $p = 0.03$.
Binning by either $\lambda = 2$ or $\lambda = 4$
attenuates the impact of bumps on the goodness-of-fit
test, which then consistently accepts the power law,
vindicating Vilfredo Pareto's 1895 assertion.

The wildfire data represents a contrasting case with
ample data recorded consistenly at $\pm$0.1 acre with
up to six digits of precision.  A basic visual
inspection of the distribution in
Fig.~\ref{fig:empirical}c reveals smooth and
substantial curvature clearly inconsistent with a power
law, even allowing for random variation. Methods for
fitting $x_m$, however, have been devised to locate
subsets of the distribution which do follow a power
law.  \citet{clauset2009power} found power-law behavior
above $x_m=6324$ using K-S statistics fitting both
$x_m$ and $\alpha$. The remaining 520 values eliminate
99.7\% of the data, indicating that the large
proportion of data does not follow a power law indeed.
Given $x_m = 6324$ however, MLEs are consistent
regardless of binning whereas the regression estimate
differs, possibly indicating curvature persisting
beyond 6324.  The $p$-values with and without binning,
however, are poles apart. The goodness of fit test to
the continuous Pareto distribution accepts the null
hypothesis with $p=0.26$, consistent with the
stipulation for choosing $x_m$ initially. Tuning the
bin width, $\lambda$, however, drives the
goodness-of-fit test towards rejecting the power law.
For $\lambda =4$, corresponding to four bins with only
one data point in the fourth bin, we reject the power
law with $p=0.01$.  As we see in the other empirical
cases, K-S statistics for continuous data ascribes
undue weight to smoothness of the data, whereas in the
present case binning conversely causes the K-S test to
ascribe more weight to the overall shape of the
distribution. The result is that binning allows the K-S
statistics to reveal curvature in the empirical
distribution when they otherwise do not.

\section{Discussion}

On perfect power-law data, estimates and
goodness-of-fit tests give consistent results with and
without binning, and binning can only reduce
statistical power \citep{virkar2014power}. When the
data deviates from a perfect power law, however,
binning may reduce biases in inferences. Logarithmic
binning preserves a power law with the original
exponent, and so we argue that the resulting discrete
power law \citep{newberry2019self} is a better baseline
model for real data than the Pareto distribution.
When conclusions depend on binning, as we observe in
the empirical data, the data deviate from a perfect
power law, possibly for trivial reasons such as small
errors. In this case, inferences without binning cannot
be trusted, as the deviations bias the inference, and
inferences based on binned data are more robust.

We find that logarithmic binning can attenuate the
affects of additive and multiplicative noise and linear
and logarithmic quantization errors. Real data contains
many more sources of error, such as censorship or
reporting biases, various kinds of measurement error,
or dependence between samples. The argument for
logarithmic binning makes no reference to the exact
structure of the errors so long as the errors do not
tend to cause the average addition or removal of data
from particular bins. This requirement is much less
stringent than requiring samples to be within
statistical deviation of the perfect
continuous power law. For errors such as noise and
quantization that have small effects on individual
measurements, the requirements can be assured if the
binning ratio $\lambda$ is sufficiently large relative
to the scale of the error. How large is sufficient
depends on the purpose of the inference and the size of
the sample (Figs.~\ref{fig:lambda}, \ref{fig:sigman},
and \ref{fig:alpha}).

Loose specification of errors is convenient in
practice, because the details of error sources are
often unknown. Nonetheless, we expect that binning is
effective in attenuating some error types more than
others. Binning may be less effective against errors
that cause deviation from the pure power law over its
full range, such as scale-dependent
censorship \citep{woessner2005assessing}, contamination,
or dependence between samples
\citep{gerlach2019testing}. These cases need further
investigation, but our study nonetheless offers some
clues.

Model misspecification near $x_m$ and log-periodic
fluctuations are two common causes of deviation from
pure power laws. We do not study these cases
explicitly, but our results do have bearing.
Often, values near $x_m$ do not follow a perfect
power law and log-log tail distribution plots show
substantial curvature in small values. Small values may
be censored, as in earthquake reporting, or may represent a bulk
distribution with distinct behavior from the tail.
Despite dealing only with small measurement noise, our
validation procedure is nonetheless a reasonable proxy
for both cases. Because we simulate noise by sampling
from a power law above $x_m$, applying error, and then,
of necessity, fitting using only the data above $x_m$,
the procedure acts both to create a censorship
process---because data reduced below $x_m$
by noise are removed from the sample---and a bulk
distribution distinct from its tail---because applying
normal or lognormal error to power law samples
yields a unimodal distribution peaked near $x_m$. Thus
samples with higher error such as those in
Fig.~\ref{fig:sigman} show more of the notorious
curvature for small values common in empirical data. In such
cases, increasing $x_m$ is known to reduce biases
\citep{danielsson2001using}. Our results show that increasing $\lambda$
also reduces bias from this type of error, offering an
alternative or additional measure of conservatism
besides increasing $x_m$.

Log-periodic fluctuations are another deviation from
power-law behavior common in empirical data. This
phenomenon can result from dependence between data
points in the sample, such as large blood vessels
branching into smaller
vessels \citep{newberry2015testing} or large social
groups composed of proportionally smaller
units \citep{zhou2005discrete}. While Pareto samples
display a continuous scale invariance, data with
log-periodic fluctuations has a discrete-scale
invariance, with a fundamental scaling ratio $\lambda$
set by the ``wavelength'' of the fluctuations
\citep{sornette1998discrete}. If the bin width $\lambda$
matches the scaling ratio $\lambda$, the discrete-scale
invariance of the model matches that of the data,
conformity to the discrete power law is restored and
inferences will be valid \citep{newberry2019self}.
Thus, we expect
data with log-periodic fluctuations may be accurately
fit with logarithmic binning as long as care is taken
to choose bin widths $\lambda$ that are integer powers
of the fundamental scaling ratio of the data.

Our findings challenge some conventional wisdom about
power law inference while addressing some ongoing
concerns. Maximum likelihood estimation offers strong
mathematical guarantees whereas regression has a
manifestly incorrect error model
\citep{goldstein2004problems,bauke2007parameter,clauset2009power,stumpf2012critical}.
At the same time, the fragility of maximum likelihood
in the case of power laws has been underappreciated,
and we find that appropriate regression methods are
relatively robust in practice (cf.
Fig.~\ref{fig:fits}, \citet{gabaix2011rank}).
Logarithmic binning offers MLEs the robustness of
regression while maintaining the logical footings of
maximum likelihood. The parameter $\lambda$ offers a
smooth interpolation between the ideal precision of the
continuous MLE and the accuracy of a vague and
broadly-applicable error model.

Considerable discussion in recent decades has
questioned the validity and the value of power laws
\citep{stumpf2012critical}, including studies that have
rejected putative power laws on statistical grounds
\citep{clauset2009power}. Often, however, these studies
have left zero tolerance for error
\citep{gerlach2019testing}. Binning offers a
way to re-evaluate the quality of empirical power laws
with some allowance for trivial experimental or data
collection error.

Debate about the validity of power laws has also
sometimes conflated methods to detect power laws and
methods to measure the exponent, whereas these are two
quite different scientific concerns. Tuning $\lambda$
allows empiricists to use different models  within the
same model family to answer different questions. For
example, we find that four bins preserve the maximum
ability to reject lognormal data in hypothesis tests,
whereas two bins are sufficient for parameter
inference, where a power law is necessarily assumed and
minimizing bias is the utmost concern.

Finally, estimating the sample fraction for tail
behavior or $x_m$ is an old and ongoing problem for
inference \citep{hall1985adaptive,drees2020minimum}. For
simplicity, our analysis takes $x_m = 1$ to be given,
whereas in practice, a true $x_m$ parameter is
typically unknown and may not ever perfectly separate a
distribution's bulk behavior from its tail. Model-free
approaches to estimating $x_m$ have been developed for
Pareto tails \citep{clauset2009power} and binned
power-law tails \citep{virkar2014power}.
Underestimating $x_m$, however introduces biases into
estimates of $\alpha$ by contamination from the bulk.
Binning offers no solution to accurately choosing
$x_m$, but does attenuate error from accidental
contamination from the bulk distribution resulting from
underestimation of $x_m$.

We conclude that logarithmic binning---combined with
appropriate maximum likelihood estimators and
goodness-of-fit tests---offers a rough but effective
control for the effects of common data errors on
power-law inference. These errors otherwise make
power-law inference unreliable.  We call for better
methods supporting robust inference in the many
scientific contexts in which power laws arise. Given
the ubiquity of power laws in nature and errors in
data, we hope that the methods we describe here will be
widely adopted.

\end{document}